\documentclass[12pt]{article}
\usepackage{epsfig}
\begin{document}

\centerline{\Large P.Grinevich, A.Mironov, S.Novikov}

\vspace{1cm}

\centerline{\bf New Reductions and Nonlinear Systems for the 2D
Schrodinger Operators}

\vspace{1cm}

{\bf Abstract}. {\it New Completely Integrable $(2+1)$-System is
studied. It is based on the so-called $L-A-B$-triples $L_t=[H,L]-fL$
where $L$ is a 2D Schrodinger Operator. This approach was invented
by S.Manakov and B.Dubrovin, I.Krichever, S.Novikov(DKN) in the
works published in 1976  {\footnote {It was developed later by many
authors. Specific  case with  2nd order  Schrodinger Operator $H$
  gauge equivalent in fact to our new system over complex field,  appeared first time
   as the initial example of S.Manakov (who considered this system uninteresting,  publishing it
  under the pressure of S.Novikov for justifying the
 idea: What is a right 2D analog of Lax pairs?) No investigation
of it  was performed later, it  never was mentioned in the later
literature. No reductions, no theory of special solutions. No
applications of this specific system to the inverse spectral theory
for 2D elliptic self-adjoint operators with data borrowed from one
energy level, has been made (which was a main program of DKN started
in 1976)-- probably because  no nontrivial self-adjoint reductions
are known for this system even now. By this reason we call this
system and its reductions  GMMN-Grinevich, Mironov, Manakov,
Novikov, reserving for the reduction below--see the case III--
notation $ B_2$-system  as a ''2D Burgers''.} }. A nonstandard
reduction for the 2D Schrodinger Operator (completely different from
the one found by S.Novikov
  and A.Veselov in 1984)
 compatible with time dynamics of the  new Nonlinear System, is studied here.
 It can be naturally treated as a 2D extension of the famous
  Burgers System. The Algebro-Geometric (AG) Periodic Solutions here are very specific and unusual  (for
  general and reduced cases). The reduced system is  linearizable like Burgers.
   However,   the general one  (and probably the reduced one also)
certainly
   lead in the stationary AG case to
  the nonstandard examples of algebraic  curves $\Gamma\subset W$   in the full complex 2D manifold
  of Bloch-Floquet functions $W$ for the periodic elliptic 2D  operator $H$ where
  $H\psi(x,y,P)=\lambda(P)\psi(x,y,P),P\in \Gamma$. However, in the nontrivial cases
  the operators are nonselfadjoint. A Conjecture is formulated that for the nontrivial selfadjoint elliptic 2D
  Schrodinger operators $H$  with periodic coefficients {\bf The Whole 2D Complex Manifolds $W$
  cannot not contain any Zariski open part of algebraic curve $\Gamma$ except maybe one selected  level
  $H\psi=const$ found in  1976 by DKN}.
  This version contains new results. It also corrects some  non-accurate claims; in particular, the non-reduced
  system is non-linearizable in any trivial sense.}

\vspace{1cm}

{\bf 1. 2D Schrodinger Operators and Nonlinear Systems. The new system.}

 As it was pointed out in the work \cite{M} and developed in $\cite{DKN}$,
the right $2+1$ analog of $1+1$ Lax Pairs $L_t=[H,L]$  for the
nontrivial second order Schrodinger Operators
$L=\Delta+G'\partial_x+G\partial_y+S$ is $$L_t=[H,L]-fL$$ {\bf We
always will use gauge condition $G'=0$}. We call them {\bf
$L-A-B$-triples where $A=H$ and $B=f$ in this notations}.
 Such equation is equivalent to the equations $$(L_t-[H,L])\psi=0$$ for all $\psi$ such that $L\psi=0$.
 A lot of works was written by the Moscow Soliton group about these systems and {\bf Inverse Spectral (Scattering )
 Problems for the elliptic 2D Schrodinger operators $L$ with periodic and rapidly decreasing coefficients based on the data
 collected from  One Energy Level only}.
(see \cite{M,DKN,KN,Ch,NV1,NV2,NV3,KNV,GRoN,G,GM,GN,G2,DKN2}).
Special attention was paid  to the {\bf Purely Potential Reduction}
$G=0$. It was found effectively in \cite{NV1,NV2} and leads to the
Prym's $\Theta$-functions in the periodic case. Corresponding
Nonlinear Systems (the NV-Hierarchy) always have operators $H$ of
odd order in each variable. In particular, the first nontrivial
operators $H$ have order 3 in both variables $x,y$ and present a 2D
extension of the famous KdV system different from KP.

We consider in this work the system $L_t=[H,L]-fL$ where
$$L=\partial_x\partial_y+G\partial_y+S, H=\partial_x^2+\partial_y^2
+F\partial_y+A$$

 {\bf Proposition 1}. {\it Following GMMN system of
evolutional equations follows from the $(L-H-f)$-triple:
$$G_t=G_{xx}-G_{yy}+(F^2/4)_x-(G^2)_x-A_x+2S_y$$
$$S_t=S_{yy}-S_{xx}-2(GS)_x+(FS)_y$$
with differential constraints $$F_x=2G_y, A_y=2S_x, f=2G_x-F_y$$ The
reduction $S=0$ is well-defined. It is a special 2D extension of the
famous Burgers system. So we call it $B_2$.}

{\bf Proposition 2}. {\it The elementary substitution:
$$G=-(\log c)_x,F=-2(\log c)_y$$
$$A=-2u_x, S=-u_y$$ leads to the form $$[c^{-1}(c_t-c_{xx}+c_{yy})]_x=-2(u_{xx}-u_{yy})$$
$$(u_t+u_{xx}-u_{yy}+2u_yc_y/c)_y=2(u_yc_x/c)_x$$ The condition $S=u_y=0$ is time invariant, and
the $B_2$ system reduces to the linearized form:
$$c_t-c_{xx}+c_{yy}=(\Phi(y,t)+A(x,t))c$$ So the
 reduced system $B_2$ is  linearizable in the variable $c$ as the ordinary Burgers corresponding to the
 $y$ independent solutions of the $B_2$ system.}

   Both statements can be easily
checked by  direct calculation.

{\bf Corollary 1.}{\it Every product function
$c_i=c'_i(x,t)c''_i(y,t)$ satisfying to the pair of separate
equations in the variables $x$ and $y$
$$c'_{it}=c'_{ixx}+A(x,t)c'_i,c''_{it}=-c''_{iyy}+\Phi(y,t)c''_{iyy}$$
satisfies to the full equation. Every linear combination of such
solutions $c=\sum_ic'_i(x,t)c''_i(y,t)$ satisfies to the full
equation for $c$ leading to the family of  coefficients $F=-2c_x/c$
and $G=-c_y/c$ nontrivially depending of all variables.}

{\bf How to describe effectively the reduction
 $S=0$ for the Bloch functions of the AG 2D Schrodinger Operator $L=\partial_x\partial_y+G\partial_y+S
 $? How to find Algebro-Geometric Solutions to the linear equation $L\psi=0$
 and nonlinear equation $L_t=[H,L]-fL$?}

\vspace{1cm}

{\bf 2. The Algebro-Geometric Solutions}.

Let us start with nonsingular Algebraic Curve $\Gamma$ accompanied by the standard Inverse Spectral Data consisting of 2
 infinite points $P_1,P_2$ with local parameters $z_1=1/k_1,z_2=1/k_2$ correspondingly, and with selected generic
 set of $g$ points
 $D=Q_1+...+Q_g$. We construct a function $\psi(x,y,t,P)$ on $\Gamma$ depending on parameters $x,y,t$ such that:

 1.$\psi$ is meromorphic on $\Gamma$ except the points $P_1,P_2$ where it has asymptotic
 $$P_1: \psi=c_1(x,y,t)\exp\{k_1x+k_1^2t\}(1+u/k_1+...)$$
$$P_2:\psi=c_2(x,y,t)\exp\{k_2y+k_2^2t\}(1+a/k_2+...)$$

 2.$\psi$ has exactly $g$ poles of the first order whose position
 is independent of parameters $x,y,t$.

  3.$\psi$ should be normalized. We  use  the standard normalization condition $\psi(0,0,0,P)=1$ and $c_1=1$
for the general  operator $L$ with gauge condition  $G'=0$ (here may
be $S\neq 0$).
 It is well-known that in order to choose another gauge condition $S=0$ for
 the general case, i.e.
$L=\partial_x\partial_y+G'\partial_x+G\partial_y$,   we need to use
the ''condition $K_0$'': $\psi(x,y,t,P_0)=1$
 fixing any arbitrary selected point $P_0\in \Gamma$ distinct from $D,P_1,P_2$.

 An extension of  this approach will be used for the most important case III containing the main result
  of the present work.
 We are going to use in the case III below such normalization of $\psi$ that   additional $k$ points
 and value of $\psi$ in all of them are fixed (we call it ''the condition $K_k$'').
 In this case we need to take divisor $D$
 of the degree $g+k$. Such trick was used by Krichever few years ago in the works dedicated to orthogonal
 coordinates. We will modify
  it below and use for our goals. It has some weak points (f.i. it is noneffective in terms of formulas) but
  its
  version
   is absolutely necessary for us.

As usually, we obtain 2 following statements:

{\bf Statement 1.} {\it The functions $\psi$ satisfy to the
equations
$$(\partial_t-H)\psi=0, L\psi=0$$ with $P$-independent coefficients
$$c=c_2,c_1=1,G=-c_x/c,S=-u_y=-a_x+G_y , F=-2c_y/c $$
$$A=-2u_x=-2a_y+2c_y^2/c^2+c_t/c-\Delta c/c
$$}

{\bf Statement 2.} {\bf Let an algebraic function $\lambda(P)$
exists on $\Gamma$ such that it has only 2 poles of second order in
the points $P_1,P_2$ with principal parts equal exactly to
$k_1^2,k_2^2$. Then $H\psi=\lambda\psi$ for all $P\in\Gamma$.}

Anyway, the compatibility requirement leads to the nonlinear
equation above $L_t=[H,L]-fH$. In the second case we have $L_t=0$.

There are following cases:

{\bf Case I}: No restriction, Standard Normalization, Nonsingular
Riemann Surface. We obtain the algebro-geometric solutions to the
generic system above. Their reality condition (all coefficients
$F,G,S,A$ are real)
 can be easily found:

 {\bf Lemma 1.} {\it 1. Let  nonsingular Riemann Surfaces of finite genus with antiholomorphic
involution $\tau$ be given  such that
$$\tau(P_j)=P_j,\tau^*(k_j)=-\bar{k}_j,\tau(D)=D$$
Then all coefficients $F,G,S,A$ are real. 2. Let also an algebraic
function $\lambda$ be given such that it has only second order poles
in the points $P_1,P_2$ with principal parts $k_1^2,k_2^2$. Than
$H\psi=\lambda\psi$ for the real elliptic second order operator
$H=\Delta+F\partial_y+A$}

{\bf Another interesting class is such that the operator $H$ is
self-adjoined. Is it possible?} Formally, such class can be obtained
from the same class of ''real'' Riemann surfaces and antiholomorphic
involutions but we should take divisors $D$ satisfying to the
equation
$$D+\tau D\sim K+ P_1+P_2$$ (we call it ''The Cherednik Type Equation'').
{\bf However, this equation is not solvable for the nontrivial
Riemann Surfaces. It is solvable for such cases that operator $H$
has odd orders in both variables $x,y$. For the second order $H$ it
leads to the trivial class only.}

Easy to see directly from the nonlinear equations that ''physical''
initial values are not invariant under the time dynamics. Here
$iF_x/2\in R$ and $eU=A-F_y/2-F^2/4\in R$ (they are called Magnetic
Field $eB$ and Electric Potential $eU$ by physicists). Their reality
is necessary for the operator $H$ to be self-adjoint.

{\bf Conjecture. Consider  any nontrivial self-adjoint smooth
periodic (topologically trivial) second order operator
$$H=\partial_x^2+(\partial_y+F/2)^2+U=\Delta+F\partial_y+A$$
$$A=U-F^2/4-F_y/2$$
(i.e. $F$ and $A$ both are periodic in $x$ and in $y$ with periods
$(T_1,T_2$). Construct its full 2D complex manifold $W$ of all
Bloch-Floquet Formal Eigenfunctions $\psi(x,y,P)$ :
$$\psi(x+T_1,y,P)=\mu_1\psi(x,y,P),\psi(x,y+T_2,P)=\mu_2\psi(x,y,P)$$
$$H\psi=\lambda(P)\psi,P\in W$$
Let an algebraic nonsingular curve $\Gamma$ be given such that its
Zariski open part $\Gamma^*$ is a complex submanifold in
$W$,$$\Gamma^*\subset W, \Gamma=\Gamma^*\bigcup \infty$$ Here
$\infty$ is a finite set of points. Than either  the surface
$\Gamma^*$ coincides with some ''energy'' level $\lambda=const$ or
coefficients of operator $H$ can be reduced to the sums of functions
of one variable.}

Let us remind here that  Feldman, Trubowitz and Knorrer proved in
1990s (see \cite{FKT}) a ''Novikov Conjecture'' (formulated in early
1980s): For any real smooth periodic purely potential 2D Schrodinger
Operator 2D its Complex Bloch-Floquet manifold $W$ of eigenfunctions
can be a Zariski open part of algebraic variety  only if  potential
is equal to the sum of functions of one variable,
$$A=U=f(x')+g(y')$$ Here $x',y'$ are orthonormal coordinates.
Probably, even  stronger statement is true formulated above and
confirmed by the algebro-geometric examples.

 {\bf Case II}: Take purely potential NV-reduction
on $L$ where $c_1=c_2=c=1$. Easy to prove that in this case we have
$F=G=0, S=g(x+y)+h(x-y)$, so this case leads to the trivial result:
our new system is compatible with this reduction only for trivial
operators. From the nonlinear system we deduce immediately that
$$A_x=2S_y, A_y=2S_x, F_x=2G_y, F_y=2G_x, f=0$$

In the stationary case we need to have algebraic function $\lambda$
with 2 poles only in the points $P_1,P_2$ with principal parts
$k_1^2,k_2^2$ and holomorphic involution
$$\sigma:\Gamma\rightarrow\Gamma, \sigma^2=0$$
But we need also to have divisors $D$ such that $D+\sigma(D)\sim
K+P_1+P_2$. Unfortunately, such divisors do not exist in the
nontrivial cases where potential $A\neq f(x-y)+g(x+y)$ for the
second order purely potential operators $H=\Delta+A$.

For that we need operators $H$ of the odd orders in both variables
$x,y$. The so-called NV-systems are always such that both these
orders are equal to each other. Let us mention that  Taimanov in
\cite{T} and recently Krichever in \cite {Kr} considered the whole
hierarchy where orders can be different (for the investigations of
''Novikov Conjecture'' for the analog of Riemann-Shottki Problem for
the Prym's $\Theta$-functions).

 {\bf Case
III}: Consider now our new reduction $S=0$ (see above: we have
$A(x,t)$ non necessary equal to zero here). How to find
algebro-geometric solutions? Let us describe here spectral data
corresponding to this case. Our picture can be viewed as a result of
degeneration in the family of algebraic curves $\Gamma_{\epsilon}$
for $\epsilon\rightarrow 0$. In the final moment we obtain a group
of transversal crossings to the points where small cycles
degenerated. In our specific case the collection of vanishing cycles
divides Riemann Surface $\Gamma$ into 2 pieces
$\Gamma=\Gamma'\bigcup \Gamma''$.
\begin{center}
\mbox{\epsfxsize=12cm \epsffile{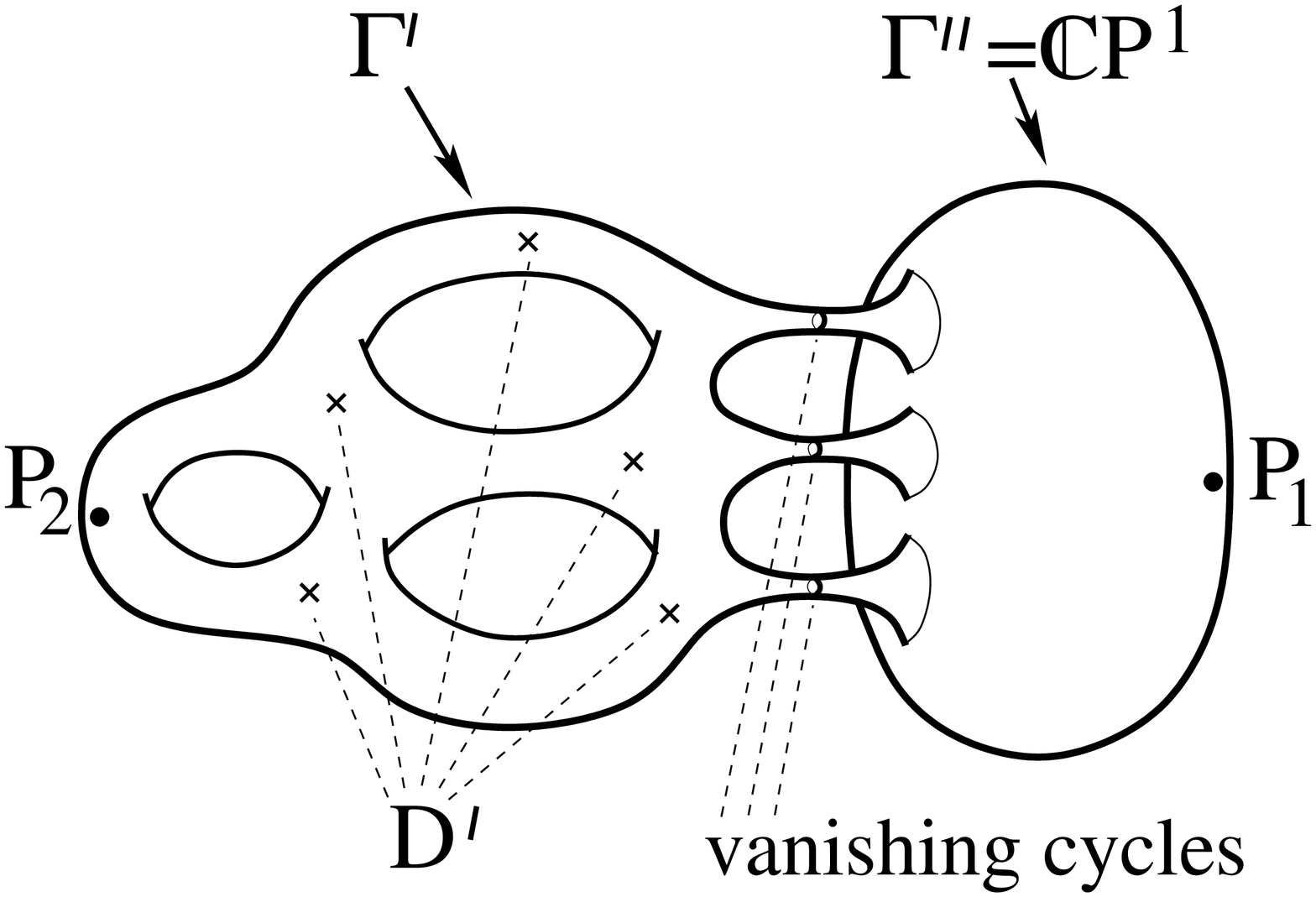}}

Fig 1. The case $A=0$.
\end{center}
 One of them $\Gamma''$ should have
genus equal to zero in the case $A=const$. If $A(x)$ does not depend
on $t$, we require it to be some finite-gap potential from KdV
theory $w=A'(x)$, and $\Gamma''$ should be corresponding
hyperelliptic Riemann surface with selected point $P_1\in \Gamma''$.
\begin{center}
\mbox{\epsfxsize=12cm \epsffile{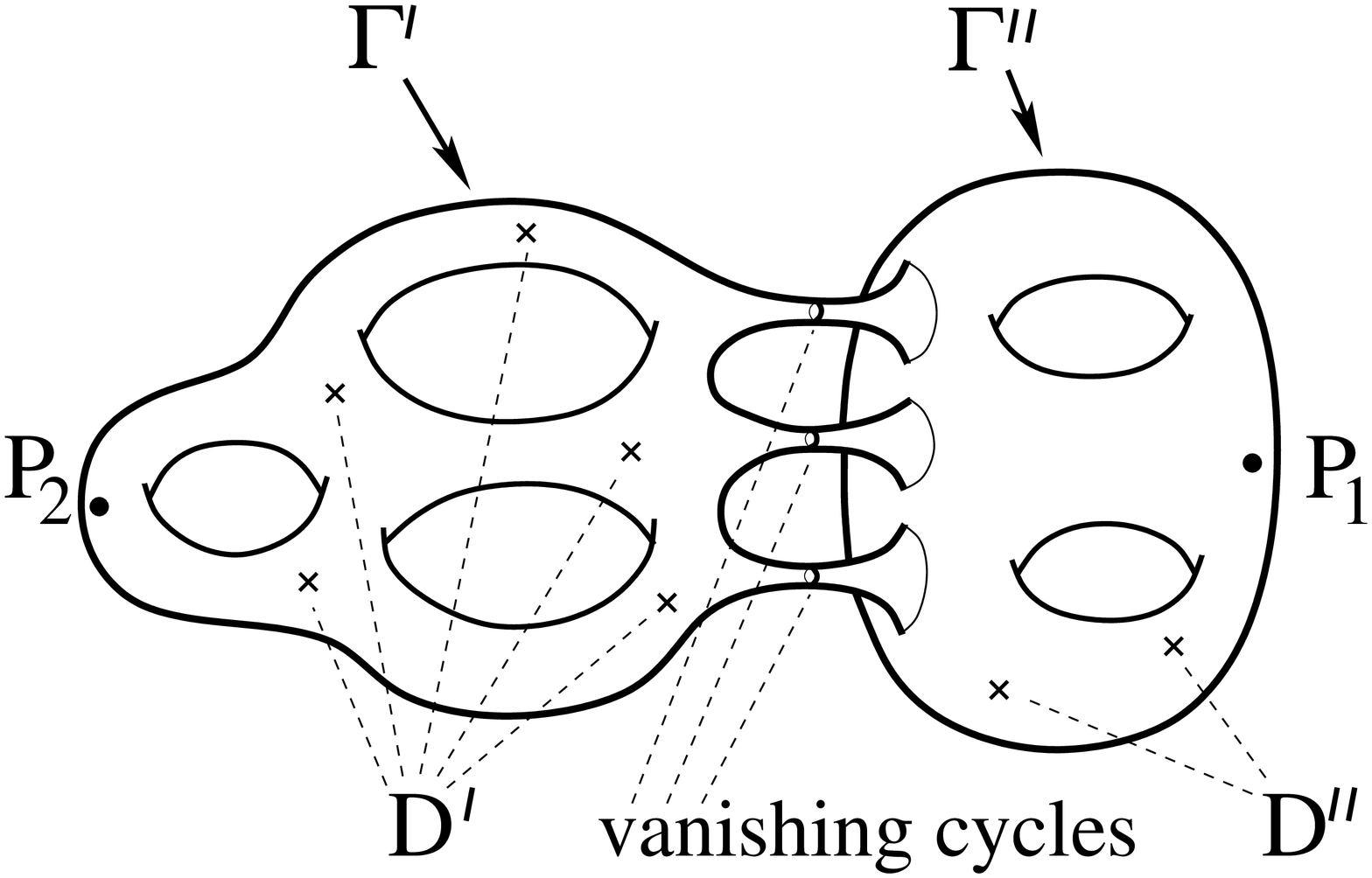}}

Fig 2. The case $A\ne 0$.
\end{center}
It is equal to the Weierstrass point with corresponding canonical
local parameter. For the general finite-gap (AG) time-dependent (KP)
case we take any nonsingular Riemann surface $\Gamma''$ with
selected point $P_1$ and local parameter $k_1$. We construct as
usual  a Baker-Akhiezer Krichever type function $\psi''(x,t,P)$ on
$\Gamma''$ with divisor $D''$ and asimptotic
$$\exp\{k_1x+k_1^2t\}(1+...)$$ near $P_1$.

So we finally need to take degenerate Riemann surface $\Gamma$ such
that
$$\Gamma=\Gamma'' \bigcup \Gamma', P_1=\infty\in \Gamma'',  P_2\in \Gamma'$$
Let genuses of these surfaces be $g',g''$ correspondingly. The
intersection of them is a discrete set outside of infinities
$P_1,P_2$. We assume that it is finite. The Riemann Surface
$\Gamma'$ should be also algebraic. The union of this assumptions is
called a ''Finite-Gap Property'' in our case. So we have:
$$\Gamma''\bigcap \Gamma'=R=(R_0\bigcup R_1\bigcup...\bigcup R_k)$$ Here
$R$ is a union of $k+1$ distinct points $R_q$ imbedded  into the
both Riemann surfaces $R\rightarrow R'\subset \Gamma'=S^2$ and
$R\rightarrow R''\subset \Gamma''$. We assume that both $R',R''$ do
not meet infinite points $P_2,P_1$ and divisors,$D',D''$. The
divisor $D$ is completely concentrated in the part $\Gamma'$ for the
case $A=0,\Gamma''=CP^1$ where $D''$ is empty: there are no divisor
points in $CP^1=\Gamma''$. For the general case we have $D=D'\bigcup
D''$ correspondingly in $\Gamma',\Gamma''$ of the degrees $g'+k$ for
$D'\subset \Gamma'$ and $g''$ for $D''\subset \Gamma''$.

Let us point out that the points $R'_s\in CP^1$ can be identified
with the wave numbers $p_s$ such that $p_0=0$. The simplest natural
functions in this case are:
$$\phi_s(x,t)=\exp\{p_sx+p_s^2t\}, \phi_0=1$$
In the case $A\neq 0$ these points in $\Gamma''$ define wave
functions $\psi''(x,t,R''_s)=\phi_s(x,t)
$

{\bf Remark.}{\it We can take any solutions $\phi_s(x,t)$ to the
equation
$$(\partial_t-\partial_x^2)\phi_s(x,t)=0, s=1,...,k$$ instead of simple
exponents for the case $\Gamma''=CP^1$. We always require
$\phi_0=1$. For the case $A\neq 0$ all
our functions $\phi_s(x,t)$
satisfy to the equation
$$(\partial_t-\partial_x^2-A(x,t))\phi_s=0$$}

{\bf Theorem 1.} {\it Let $\Gamma'$ be a nonsingular algebraic
curve, the point $P_2,P_1$ and divisor $D'$ of degree $g'+k$ is
generic. There exists an unique function $\psi'(x,y,t,P\in\Gamma')$
with following properties: It has exponential asymptotic near the
point $P=P_2\in\Gamma'$ as above $$
k=k_2,\psi=c(x,y,t)\exp\{ky+k^2t\}(1+v(x,y,t)/k+...)$$ It has first
order poles in the fixed points $D'_s,s=1,2,...,d+k$,
and$$\psi'(x,y,t,R_s'')=\phi_s(x,t),s=0,1,...,k$$ This functions
satisfy to the equations (below) for all $P\in\Gamma'$
$$(\partial_t-H)\psi=0,L\psi=0$$ where $$G_y-v_x=S=u_y=0,A=-u_x=A(x,t),  G=-c_x/c,F_x=2G_y, F=-2c_y/c$$
$$G_t=(\partial_x^2-\partial_y^2)G + (F^2/4-G^2)_x$$}

The proof is more or less standard. Except cancelation at
infinities, we get also cancelation of values in all selected points
which is possible only if $S=0, A=A(x,t)$ . It essentially implies
our result unifying this property with the traditional arguments for
this area. The important breakthrough for us was to find these data.

 {\bf Corollary 2. Let a nonsingular Riemann
surface $\Gamma'$ be such that there exists a meromorphic function
$\lambda(P)$ with only pole of second order in $P_2\in \Gamma'$. It
means that this surface is hyperelliptic $\mu^2=P_{2g+1}(\lambda),
P_2=\infty$. Then coefficients of both operators do not depend on
$t$ and
$$H\psi=\lambda\psi(x,y,P), H=\Delta+F(x,y)\partial_y$$}
{\bf Can these operators be trivial if we take arbitrary data?}

 In
the next work we will clarify all this details and present more
precise formulas. There is something  strange in this linearizable
''Burgers Style'' Reduction $S=0$: {\it Does it generate nontrivial
examples for the Spectral Theory according to the present results or
in fact these examples are in some sense trivial? We constructed
them above but did not investigated them enough to prove this
statement completely. We still have some doubts.}

 {\bf The  reduced equation can be viewed as a
special Integrable 2D Extension of the famous Burgers Equation. The
reduced system is linearizable, the full system is non-linearizable.
So many solutions of it can be found completely elementary, in style
of the ordinary Burgers equation. However, the stationary solutions
for the AG cases possibly leads to the nontrivial examples of the
Algebraic Curves $\Gamma'\in W$ such that $H\psi=\lambda\psi$ and
$\lambda\neq const$. We know that for the full system $S\neq 0$
these type of
 examples certainly are nontrivial. Some of them for $g=2$
 were extracted by the authors from the analysis (made by
Mironov in \cite{Mir}) of the 2D matrix operators of Sato and
Nakayashiki \cite{SN} with algebraic 2D Bloch-Floquet manifold $W$.
They are non-selfadjoint, but elliptic, and data can be easily
chosen to make them real and nonsingular periodic}.

\end{document}